\documentclass[aps,showpacs,prb,reprint]{revtex4-1}

\usepackage[colorlinks=true,linkcolor=blue,citecolor=blue,urlcolor=blue]{hyperref}
\usepackage{setspace} 
\usepackage{graphicx}
\usepackage{amsmath}
\usepackage{color}
\usepackage{amsmath}
\usepackage{amssymb}
\usepackage{verbatim}
\usepackage{latexsym}
\usepackage{enumerate} 
\usepackage{bm} 
\usepackage{longtable}

\begin{document}

\title{Theory of temperature dependent phonon-renormalized properties}

\author{Bartomeu Monserrat}
\email{bm418@cam.ac.uk}
\affiliation{TCM Group, Cavendish Laboratory, University of Cambridge, J.J. Thomson Avenue, Cambridge CB3 0HE, United Kingdom}

\author{G.J. Conduit}
\affiliation{TCM Group, Cavendish Laboratory, University of Cambridge, J.J. Thomson Avenue, Cambridge CB3 0HE, United Kingdom}

\author{R.J. Needs}
\affiliation{TCM Group, Cavendish Laboratory, University of Cambridge, J.J. Thomson Avenue, Cambridge CB3 0HE, United Kingdom}

\date{\today}

\begin{abstract}

We present a general harmonic theory for the temperature dependence of phonon-renormalized properties of solids. Firstly, we formulate a perturbation theory in phonon-phonon interactions to calculate the phonon renormalization of physical quantities. Secondly, we propose two new schemes for extrapolating phonon zero-point corrections from temperature dependent data that improve the accuracy by an order of magnitude compared to previous approaches. Finally, we consider the low-temperature limit of the class of observables that includes the electronic band gap, obtaining a $T^4$ dependence in three dimensions, $T^2$ in two dimensions, and $T^{3/2}$ in one dimension.
\end{abstract}


\maketitle

\section{Introduction} 

To understand the physical properties of a material it is crucial to have a full description of the interactions between the electrons and atoms. The vibrations of the atoms encapsulated as phonons have a substantial impact on the properties of a material. The zero-point (ZP) phonon motion often has a significant effect, whereas the temperature dependence of several key properties, including electronic band gaps and equilibrium volumes, are driven mainly by the atomic vibrations.


The importance of considering the vibrational state of a solid was evident from the early days of quantum theory, motivating the Einstein\cite{ANDP:ANDP19063270110} and Debye\cite{ANDP:ANDP19123441404} models for the specific heat. 
We will focus on the description of the phonon-driven temperature dependence of quantities such as band gaps and thermal expansion.\cite{0295-5075-10-6-011,RevModPhys.77.1173,PhysRevB.71.205214,nanotube_t_dependence,PhysRevLett.105.265501,PhysRevB.87.144302}
First-principles calculations have revolutionized the analysis of phonons in solids\cite{PhysRevB.43.7231,RevModPhys.73.515} and are invaluable for quantitative calculations of the ZP renormalization and the temperature dependent properties of solids. In particular, the formalism proposed in Ref.\ \onlinecite{PhysRevB.87.144302} delivers accurate quantitative results that serve as a platform for a general description of phonon renormalization. We can gain further insights through a phenomenological approach by considering the general properties of the theory without reference to an underlying microscopic theory or a specific material. 
The comparison between analytic and first-principles methods leads to a wider picture of the effects of the vibrational state on physical observables.

In Sec.\ \ref{sec:theory} we present a general harmonic theory of the temperature dependence of phonon-renormalized properties of solids. 
The theory exposes important approximations, and in Sec.\ \ref{sec:perturb} we assess the accuracy of several perturbative theories.\cite{0022-3719-9-12-013,doi:10.1080/00018738000101426} In Sec.\ \ref{sec:extrap} we propose two new models for use within an extrapolation scheme for obtaining ZP corrections to quantities such as band gaps or lattice parameters from experimental data\cite{doi:10.1080/01418639408240227,Cardona20053} that are an order of magnitude more accurate than previous models. 
In Section\ \ref{sec:limits} we describe the asymptotic behaviour of band gaps at low temperatures, recovering the standard $T^4$ power law for three-dimensional systems\cite{PhysRevLett.92.196403} that arises from the linear dispersion of the acoustic branches. 
Two-dimensional systems obey a $T^2$ power law, and one-dimensional systems follow a $T^{3/2}$ power law, both dominated by the quadratic acoustic branches. We summarize our findings in Sec.\ \ref{sec:conclusions}.

\section{Mathematical formulation} \label{sec:theory}

We first construct a general framework for calculating the expectation value of an observable $O$ that depends on the vibrational state of the solid. This will be done both at perturbative and non-perturbative levels of approximation, summarized in Table\ \ref{tab:approximations}, that offer compromises between exactness, ease of calculation, and physical insight. First-principles calculations will be used in Sec.\ \ref{sec:perturb} to compare the accuracy of the different approaches, and new models for obtaining ZP corrections from experimental data will be developed in Sec.\ \ref{sec:extrap}.
Throughout this paper the phrase ``ZP correction'' refers to the correction of a general physical observable and is not restricted to the specific correction of the vibrational energy.

\newlength{\stdgap}
 \setlength{\stdgap}{5.5pt}
\begin{table*}[htbp]
\caption{Schemes for calculating phonon expectation values.}
\label{tab:approximations}
\begin{tabular}{lll}
\hline
\hline
 \textbf{Method}  & \hspace{0.1cm} \textbf{Advantages} &\hspace{0.1cm} \textbf{Disadvantages} \\
\hline
Non-perturbative &\hspace{0.1cm} Numerically exact &\hspace{0.1cm} New calculation required at each $T$ \\ 
[\stdgap]
Phonon interaction expansion &\hspace{0.1cm} Single calculation required for all $T$ &\hspace{0.1cm} Perturbative in phonon-phonon interactions \\
                                 &\hspace{0.1cm} Access to the underlying physics &\\
[\stdgap]
\hline
\hline
\end{tabular}
\end{table*}

We model a solid of $N$ atoms by a supercell subject to periodic boundary conditions. The vibrational motion of the atoms can be described within the harmonic approximation in terms of $3N$ phonon coordinates $\{q_{n\mathbf{k}}\}$, where $n$ is the branch index and $\mathbf{k}$ is a reciprocal space vector within the first Brillouin zone (BZ). In terms of phonon coordinates, the vibrational Hamiltonian $\hat{\mathcal{H}}$ reads
\begin{align}
\hat{\mathcal{H}}=\sum_{n,\mathbf{k}}\left(-\frac{1}{2}\frac{\partial^2}{\partial q_{n\mathbf{k}}^2} + \frac{1}{2}\omega^2_{n\mathbf{k}}q_{n\mathbf{k}}^2\right)\,,
\end{align}
where $\omega_{n\mathbf{k}}$ are the phonon frequencies. The energy associated with a phonon mode $(n,\mathbf{k})$ in state $m$ is $E_{n\mathbf{k};m}=\omega_{n\mathbf{k}}\left(m+1/2\right)$, and the corresponding state is $|\phi_m(q_{n\mathbf{k}})\rangle$.
We label the vibrational state of the solid by the $3N$-dimensional vector $\mathbf{M}$, whose element $M_{n\mathbf{k}}$ labels the state 
of phonon $(n,\mathbf{k})$.
All equations are given in Hartree atomic units, $\hbar=|e|=m_{\mathrm{e}}=4\pi \epsilon_0=1$.


\subsection{Non-perturbative} \label{subsec:nonperturbative}

Let $\mathbf{Q}$ be a collective phonon coordinate with elements $q_{n\mathbf{k}}$. The expectation value at inverse temperature $\beta=1/k_{\mathrm{B}}T$ with respect to the vibrational state $|\Phi_{\mathbf{M}}\rangle=\prod_{n,\mathbf{k}}|\phi_{M_{n\mathbf{k}}}(q_{n\mathbf{k}})\rangle$ is
\begin{align}
\langle\hat{O}\rangle = \frac{1}{\mathcal{Z}}\sum_{\mathbf{M}}\langle\Phi_{\mathbf{M}}(\mathbf{Q})|\hat{O}(\mathbf{Q})|\Phi_{\mathbf{M}}(\mathbf{Q})\rangle \mathrm{e}^{-\beta E_{\mathbf{M}}}\,, \label{eq:expval}
\end{align}
where $\mathcal{Z}=\sum_{\mathbf{M}}\mathrm{e}^{-\beta E_{\mathbf{M}}}$ is the partition function. We regularize the operator $\hat{O}(\mathbf{Q})$ by subtracting the static lattice value $\hat{O}(\mathbf{0})$ to focus on the correction due to the vibrational state.

This expectation value can be evaluated directly by Monte Carlo sampling weighted by the phonon density.\cite{PhysRevB.73.245202,giustino_nat_comm} Although this approach leads to numerically exact results, the random sampling obscures the underlying physical processes. The phonon density is temperature dependent so a new calculation is required at each temperature, rendering this the most computationally expensive approach.

\subsection{Phonon interaction expansion}

To gain physical insight into the dominant processes and reduce the computational expense we construct an expansion in the phonon-phonon interactions.
We first recast $\hat{O}(\mathbf{Q})$ as\cite{PhysRevB.87.144302} 
\begin{widetext}
\begin{align}
\hat{O}(\mathbf{Q})=\sum_{n,\mathbf{k}}\underbrace{a_{n\mathbf{k}}f_{n\mathbf{k}}(q_{n\mathbf{k}})}_{\hat{O}_{n\mathbf{k}}(q_{n\mathbf{k}})}+\sum_{(n,\mathbf{k})\neq(n',\mathbf{k}')}\underbrace{a_{\{n\mathbf{k}|n'\mathbf{k}'\}}f_{\{n\mathbf{k}|n'\mathbf{k}'\}}(q_{n\mathbf{k}},q_{n'\mathbf{k}'})}_{\hat{O}_{n\mathbf{k};n'\mathbf{k}'}(q_{n\mathbf{k}},q_{n'\mathbf{k}'})}+\cdots\,, \label{eq:expansion_op}
\end{align}
where $f$ is a basis set for the phonon spectrum and the set $\{a_{n\mathbf{k}},a_{\{n\mathbf{k}|n'\mathbf{k}'\}},\ldots\}$ are the coupling constants of the phonons with the observable $O$ that can be evaluated, for example, within a first-principles method. 
This allows us to rewrite the phonon expectation value as
\begin{align}
\langle\hat{O}\rangle= 
&\sum_{n,\mathbf{k}}\!\frac{1}{\mathcal{Z}_{n\mathbf{k}}}\!\sum_{m=0}^{\infty}\!\langle\phi_m(q_{n\mathbf{k}})|\hat{O}_{n\mathbf{k}}(q_{n\mathbf{k}})|\phi_m(q_{n\mathbf{k}})\rangle \mathrm{e}^{-\beta E_{n\mathbf{k};m}} \nonumber \\ 
+&\!\!\!\!\!\!\!\!\!\!\!\sum_{(n,\mathbf{k})\neq(n',\mathbf{k}')}\!\!\!\frac{1}{\mathcal{Z}_{n\mathbf{k}}\mathcal{Z}_{n'\mathbf{k}'}}\!\!\!\sum_{m,m'=0}^{\infty}\!\!\!\langle\phi_m(q_{n\mathbf{k}})\phi_{m'}(q_{n'\mathbf{k}'})|\hat{O}_{n\mathbf{k};n'\mathbf{k}'}(q_{n\mathbf{k}},q_{n'\mathbf{k}'})|\phi_{m'}(q_{n'\mathbf{k}'})\phi_m(q_{n\mathbf{k}})\rangle \mathrm{e}^{-\beta E_{n\mathbf{k};m}}\mathrm{e}^{-\beta E_{n'\mathbf{k}';m'}}\!\!+\cdots\,.
\end{align}
\end{widetext}
This perturbative method leads to numerically exact results only if sufficient phonon-phonon terms are included. However, as each phonon is treated explicitly it directly exposes the underlying physics. The most expensive stage is the first-principles computation of the coupling constants, thereafter the full temperature dependence can be studied at a small additional computational cost for systems with a band gap.  




For computational purposes a choice of basis $f$ is required in Eq.\ (\ref{eq:expansion_op}). We choose a polynomial basis $\{q^s\}$ because it leads to analytic results and connects with standard theories of thermal expansion and band gap renormalization (see Sec.\ \ref{sec:perturb} below).
Within the harmonic approximation only even functions lead to non-zero expectation values, hence the polynomial basis may be rewritten as $\{|q|^s\}$, and the relevant matrix elements are
\begin{widetext}
\begin{align}
\mathcal{M}_{s,m}=\bigl\langle\phi_m(q)\bigl||q|^s\bigr|\phi_m(q)\bigr\rangle=\frac{s!}{(4\omega)^{s/2}}\frac{2^m}{m!}\sum_{p=\big\{\substack{\max(0,m-s), \mbox{ } s\mbox{ \scriptsize{even}} \\\!\!0,\hphantom{\max(m-s)\,\,} \mbox{ }  s\mbox{ \scriptsize{odd}}}}^m\frac{({}^{m}C_{p})^2\,p!}{2^p\Gamma(s/2-m+p+1)}\,, \label{eq:matrix_element}
\end{align}
where ${}^mC_p$ is a binomial coefficient and $\Gamma$ is the gamma function. We then obtain
\begin{align}
\langle\hat{O}\rangle=&\sum_{n,\mathbf{k}}(1-\mathrm{e}^{-\beta\omega_{n\mathbf{k}}})\sum_{s=1}^{\infty}\sum_{m=0}^{\infty}a_{s;n\mathbf{k}}\mathcal{M}_{s,m}\mathrm{e}^{-m\beta\omega_{n\mathbf{k}}} \nonumber \\ 
+&\!\!\!\!\sum_{(n,\mathbf{k})\neq(n',\mathbf{k}')}\!\!\!(1-\mathrm{e}^{-\beta\omega_{n\mathbf{k}}})(1-\mathrm{e}^{-\beta\omega_{n'\mathbf{k}'}})\!\!\sum_{s,s'=1}^{\infty}\sum_{m,m'=0}^{\infty}a_{\{s;n\mathbf{k}|s';n'\mathbf{k}'\}}\mathcal{M}_{s,m}\mathcal{M}_{s',m'}\mathrm{e}^{-m\beta\omega_{n\mathbf{k}}}\mathrm{e}^{-m'\beta\omega_{n'\mathbf{k}'}}+\cdots\,, \label{eq:full} 
\end{align}
\end{widetext}
for the single and double phonon terms, and similar expressions for higher order terms. The coupling constants $a_{s;n\mathbf{k}}$ have been rewritten in terms of the polynomial basis power $s$.


Equation\ (\ref{eq:full}) describes the temperature dependence of the expectation value of observable $O$.
We use this general framework to address three questions: (i) the use of perturbation theory for the calculation of phonon-renormalized expectation values, (ii) the calculation of ZP corrections from experimental data, and (iii) the low temperature asymptote of the expectation value of a class of such observables, including the electronic band gaps. For the first two problems we validate our findings with first-principles calculations. The third problem is an example of a situation that is not directly accessible in practise to first-principles calculations due to the fine $\mathbf{k}$-point sampling required near the BZ center.

\section{Beyond lowest order perturbation theory} \label{sec:perturb}


The calculation of the phonon renormalization of many physical observables is facilitated by a perturbation expansion in the phonon-phonon interaction. We now assess the validity of these expansions by comparing the analytic results from Sec.\ \ref{sec:theory} with first-principles calculations.\cite{PhysRevB.87.144302}


To calculate the renormalization of a general observable, we start from Eq.\ (\ref{eq:full}) in the previous section. We note that if $a_s=0$ for all $s\neq 2$ in the independent phonon term, and all phonon-phonon coupling terms vanish, we obtain
\begin{align}
\langle\hat{O}\rangle=\sum_{n,\mathbf{k}}\frac{a_{2;n\mathbf{k}}}{2\omega_{n\mathbf{k}}}\left[1+2n_{\mathrm{B}}(\omega_{n\mathbf{k}})\right]\,, \label{eq:bose}
\end{align} 
where $n_{\mathrm{B}}(\omega)=(\mathrm{e}^{\beta\omega}-1)^{-1}$ is a Bose-Einstein (BE) factor. (The derivation of this result is described in Appendix~\ref{app:bose}.) This expression recovers the standard formulation\cite{doi:10.1080/01418639408240227} of the temperature dependence of band gaps within Allen-Heine-Cardona (AHC) theory\cite{0022-3719-9-12-013,PhysRevB.23.1495} and of the temperature dependence of lattice parameters within the Gr\"{u}neisen formalism.\cite{doi:10.1080/00018738000101426}
The high temperature limit of Eq.\ (\ref{eq:bose}) is 
\begin{align}
\langle\hat{O}\rangle\underset{\scriptstyle{\beta\ll1}}{\to}\left(\sum_{n,\mathbf{k}}\frac{a_{1;n\mathbf{k}}}{\omega_{n\mathbf{k}}^2}\right)\beta^{-1}\,. \label{eq:highT}
\end{align}

Many physical systems are expected to be well-described by restricting the polynomial expansion to even powers $\{q^{2t}\}$ for $s=2t$, so we first focus only on these terms. Starting from Eq.\ (\ref{eq:full}), we can systematically improve the perturbation theory beyond Eq.\ (\ref{eq:bose}). The next order term is a sum of the independent phonon term corresponding to $s=4$,
\begin{align}
\langle\hat{O}\rangle\!=\!\!\sum_{n,\mathbf{k}}\frac{a_{4;n\mathbf{k}}}{4\omega_{n\mathbf{k}}^2}\!\!\left[1\!+\!12\,\mathrm{e}^{\beta\omega_{n\mathbf{k}}}\,n^2_{\mathrm{B}}(\omega_{n\mathbf{k}})\right]\,, \label{eq:bose2}
\end{align}
and the phonon-phonon term with $s=2$ and $s'=2$, 
\begin{align}
\langle\hat{O}\rangle\!=\!\!\!\!\!\!\!\sum_{(n,\mathbf{k})\neq(n',\mathbf{k}')}\!\!\!\!\!\frac{a_{\{2;n\mathbf{k}|2;n'\mathbf{k}'\}}}{4\omega_{n\mathbf{k}}\omega_{n'\mathbf{k}'}}\!\left[1\!+\!2n_{\mathrm{B}}(\omega_{n\mathbf{k}})\right]\!\left[1\!+\!2n_{\mathrm{B}}(\omega_{n'\mathbf{k}'}\!)\right]\,. \label{eq:bose2}
\end{align}
These two perturbative terms combine to give a high-temperature limit
\begin{align}
\langle\hat{O}\rangle\underset{\scriptstyle{\beta\ll1}}{\to}\left(\sum_{n,\mathbf{k}}\frac{3a_{4;n\mathbf{k}}}{\omega_{n\mathbf{k}}^4}+\!\!\!\!\!\!\sum_{(n,\mathbf{k})\neq(n',\mathbf{k}')}\!\!\!\!\!\frac{a_{\{2;n\mathbf{k}|2;n'\mathbf{k}'\}}}{\omega_{n\mathbf{k}}^2\omega_{n'\mathbf{k}'}^2}\right)\beta^{-2}\,,
\end{align}
that dominates asymptotically over the linear term proportional to $\beta^{-1}$. More generally, the non-zero $a_s$ with the largest $s$ will dominate the high-temperature limit, giving a power law dependence of $\beta^{-s/2}$.
In general, the contributions of higher order terms beyond $q^2$ are unimportant because their coupling constants are several orders of magnitude smaller than $a_{2;n\mathbf{k}}$, justifying the widespread use of AHC theory for band gaps and the Gr\"{u}neisen formalism for thermal expansion. This means that the cross-over temperature to non-linear behaviour is high, and is irrelevant for the solid phase of the system. As an example, the cross-over temperature at which the quartic term becomes important in diamond is larger than $10^4$ K, which is beyond the melting temperature. In an experimental setting, nonlinear behaviour of the temperature dependence in the high-temperature limit could be taken as the signature of effects beyond the lowest order theory. 

We have implemented the three methods of AHC theory Eq.\
(\ref{eq:bose}), the independent phonon term in Eq.\ (\ref{eq:full}),
and the non-perturbative approach in Sec.\
\ref{subsec:nonperturbative}, for calculating the temperature
dependence of band gaps within first-principles calculations.\cite{PhysRevB.87.144302}
We have studied diamond and helium using plane-wave density functional
theory\cite{PhysRev.136.B864,PhysRev.140.A1133} with ultrasoft
pseudopotentials\cite{PhysRevB.41.7892} as implemented in the
\textsc{castep} code\cite{CASTEP}.
All calculations used supercells containing $54$ atoms, and
all energy differences were converged to within $10^{-4}$ eV per unit
cell and all stresses were converged to within $10^{-2}$ GPa. Table\
\ref{tab:helium} shows the ZP correction to the thermal band gap of
diamond within the different approximations. AHC theory
underestimates the accurate non-perturbative result by only $0.02$ eV,
supporting its widespread use. The full independent phonon term leads
to excellent agreement with the non-perturbative approach, demonstrating that it is important to take full account of the phonon dispersion, but that higher order phonon-phonon coupling terms are unimportant in
diamond.


\begin{table}[t]
\caption{ZP correction to the electronic band gap of diamond and helium, in units of eV.}
\label{tab:helium}
\begin{tabular}{lcc}
\hline
\hline
  & \hspace{0.1cm}  Diamond \hspace{0.1cm} & Helium \hspace{0.1cm}  \\
\hline
AHC theory & $-0.44$ & $+0.26$ \\
Independent phonon term & $-0.46$ & $+0.12$  \\
Non-perturbative & $-0.46$ & $+0.40$  \\
\hline
\hline
\end{tabular}
\end{table}


An important example of behaviour beyond lowest-order perturbation theory is the ZP correction to the band gap due to electron-phonon coupling in solid helium under the terapascal pressures found in white dwarf stars.\cite{helium} It is critical to have a detailed knowledge of the band gap as this has a significant impact on our understanding of white dwarf cooling, and consequently in estimates of the age of the Universe.\cite{helium}
Table\ \ref{tab:helium} also shows the ZP correction to the band gap arising from electron-phonon coupling in solid hexagonal closed-packed (hcp) helium at a pressure of $10$ TPa. 
For helium, the comparison between the non-perturbative calculation and the various levels of perturbation theory makes explicit the limitations of the perturbative calculations for this system. AHC theory leads to a ZP correction of $+0.26$ eV, which is significantly modified by including the full independent phonon term, reducing the correction to $+0.12$ eV. This large difference is caused by a linear (rather than quadratic) dependence of the electronic band gap as a function of phonon amplitude for helium, which can be described very accurately by the odd power terms in Eq.\ (\ref{eq:matrix_element}), but not within AHC theory. Unlike the ZP correction of diamond, in the case of helium even the independent phonon term fails to recover the full non-perturbative correction of $+0.40$ eV due to phonon-phonon interactions. We note that although the AHC result seems to be in better agreement with the non-perturbative result than the independent phonon term result, this is an artifact of the poor convergence of AHC theory for this system. 

In this section we have contextualized the widely used AHC theory, and presented, as far as we are aware, the first example of its failure. However, we expect that many systems are well-described by AHC theory, and in the rest of this paper we will restrict our attention to two further questions that can be addressed within this theory.

\section{Determining the zero-point correction} \label{sec:extrap}

The experimental characterization of the vibrational state of a solid is important for understanding many physical phenomena. The ZP correction to an observable is a direct measure of the coupling between the observable and the phonons. However, this quantity cannot be measured directly in experiments because it relates to an unphysical state without nuclear vibrations. With the first-principles method proposed in Ref.\ \onlinecite{PhysRevB.87.144302} we first expose the shortcomings of the different models used to extract the ZP correction from experimental data, and second propose and assess the accuracy of two new schemes. 

With knowledge of the lowest order expression for the temperature dependence of phonon renormalised quantities, 
\begin{align}
\langle\hat{O}\rangle=\frac{1}{2}\sum_{n,\mathbf{k}}A_{n\mathbf{k}}\left[1+2n_{\mathrm{B}}(\omega_{n\mathbf{k}})\right]\,, \label{eq:bose_alt}
\end{align}
where, from Eq.\ (\ref{eq:bose}), $A_{n\mathbf{k}}=a_{1;n\mathbf{k}}/\omega_{n\mathbf{k}}$,
the ZP correction to an observable $O$ is given by
\begin{align}
\langle\hat{O}\rangle_{\mathrm{ZP}}=\frac{1}{2}\sum_{n,\mathbf{k}}A_{n\mathbf{k}}\,,
\end{align}
and can be extracted as the zero temperature linear extrapolate from the high temperature limit $\beta\ll1/\omega$.\cite{doi:10.1080/01418639408240227} 
In practical applications of the extrapolation scheme to experimental data, an approximation must be made because experiments rarely reach temperatures high enough to enter the asymptotic linear limit. One can construct an analytic model $F(T,\mathbf{A})$ for the $T$ dependence of the observable over the entire temperature range, fitted using variational parameters $\mathbf{A}$. This analytical model is then used in the extrapolation. With the newly developed first-principles method presented in Ref.\ \onlinecite{PhysRevB.87.144302} that describes the temperature dependence using Eq.\ (\ref{eq:bose}), we can for the first time assess different models $F(T,\mathbf{A})$. We propose two new schemes and test them against previous models and obtain an order-of-magnitude improvement in the accuracy of the extrapolated ZP  correction. The models considered for $F(T,\mathbf{A})$  are enumerated in Table\ \ref{tab:models}. We note that the old models were not developed specifically for the ZP extrapolation, but instead to reproduce accurately the experimental data, which is usually available only at low temperatures. This might explain some of the failures in their application to extract accurate ZP corrections.


\begin{table*}[htbp]
\caption{Analytic models for the temperature dependence of phonon-renormalized quantities.}
\label{tab:models}
\begin{tabular}{ll}
\hline
\hline
 \textbf{Model}  & $F(T,\mathbf{A})$ \hspace{0.1cm} \\
\hline
Varshni & $A_0+\frac{A_1T^2}{A_2+T}$ \\
[\stdgap]
P\"{a}ssler & $A_0+\frac{A_1A_2}{2}\left\{\left[1+\left(\frac{2T}{A_2}\right)^{A_3}\right]^{1/A_3}+1\right\}$  \\
[\stdgap]
BE  & $A_0+\frac{A_1}{\mathrm{e}^{A_2/k_{\mathrm{B}}T}-1}$ \\
[\stdgap]
Double BE  & $A_0+\frac{A_1}{\mathrm{e}^{A_2/k_{\mathrm{B}}T}-1}+\frac{A_3}{\mathrm{e}^{A_4/k_{\mathrm{B}}T}-1}$ \\
[\stdgap]
Phonon dispersion & $A_0+\frac{A_1}{\mathrm{e}^{A_2/k_{\mathrm{B}}T}-1}+\frac{\mathrm{e}^{A_2/k_{\mathrm{B}}T}(1+\mathrm{e}^{A_2/k_{\mathrm{B}}T})A_3}{2k_{\mathrm{B}}T(\mathrm{e}^{A_2/k_{\mathrm{B}}T}-1)^2}+\cdots$ \\
[\stdgap]
Two step & $A_0+\frac{A_1}{\mathrm{e}^{A_2/k_{\mathrm{B}}T}-1}$  \\
& $\omega(T_{\mathrm{max}})\!=\!p_0\!+\!\frac{p_1p_0^2}{k_{\mathrm{B}}T_{\mathrm{max}}\ln(p_0/k_{\mathrm{B}}T_{\mathrm{max}})}\!+\!\frac{p_2p_0^3}{(k_{\mathrm{B}}T_{\mathrm{max}})^2\ln(p_0/k_{\mathrm{B}}T_{\mathrm{max}})^2}$ \\
& $A(T_{\mathrm{max}})=p_3\left(p_0+\!\frac{p_1p_0^2}{k_{\mathrm{B}}T_{\mathrm{max}}}\!+\!\frac{p_2p_0^3}{(k_{\mathrm{B}}T_{\mathrm{max}})^2}\right)$ \\
\hline
\hline
\end{tabular}
\end{table*}

 A widely used model proposed by Varshni\cite{Varshni1967149} reproduces the high-temperature linear asymptote, but incorrectly assumes a $T^2$ dependence as $T\to0$. 
P\"{a}ssler\cite{Passler} proposed a more complicated expression, which describes the low temperature behaviour by a fitting parameter that in principle could recover the low temperature $T^4$ limit (see Sec.\ \ref{sec:limits} below and Ref.\ \onlinecite{PhysRevLett.92.196403}). However, the low-temperature asymptote has little impact on the high-temperature limit or the ZP correction because the cross-over between a power law and the exponential dependence of Eq.\ (\ref{eq:bose}) occurs at very low temperatures (below $4$\,K for silicon\cite{PhysRevLett.92.196403}) and, moreover, the acoustic phonons that dominate in this regime have a low density of states.
This motivates neglecting the low temperature $T^4$ power law and instead focusing on the higher energy phonon branches that can be described by the Einstein approximation. This leads to a functional form consisting of a single BE oscillator,\cite{PhysRevB.30.1979} which amounts to assuming a dispersionless phonon spectrum. 
As Eq.\ (\ref{eq:bose}) consists of a sum over many BE oscillators, a straightforward extension of the single BE oscillator model is to include a second oscillator.\cite{2be_oscillator_fit} For systems with non-monotonic temperature-dependent gaps, characterized by more than one Einstein frequency, the use of more than a single BE oscillator is essential.\cite{PhysRevB.86.195208}

\subsection{New models for the linear extrapolation scheme}

The BE oscillator model may fail to recover the ZP correction unless data exists up to high temperatures $k_{\mathrm{B}}T\gtrsim\omega$. This motivates us to propose two new methods, based on a single BE oscillator fit,\cite{PhysRevB.30.1979}
\begin{align}
F(T,\mathbf{A})=\frac{A}{\mathrm{e}^{\omega/k_{\mathrm{B}}T}-1}\,, \label{eq:model}
\end{align}
where $\mathbf{A}=(A,\omega)$. The two new models recover the correct ZP correction even with data restricted to low temperatures. 

\subsubsection{Phonon dispersion method}

We start from Eq.\ (\ref{eq:bose_alt}), rewrite the phonon dispersion as $\omega_{n\mathbf{k}}=\overline{\omega}+\delta_{n\mathbf{k}}$, and retain the relevant temperature dependent terms, so that
\begin{align}
\sum_{n\mathbf{k}}\frac{A_{n\mathbf{k}}}{\mathrm{e}^{\omega_{n\mathbf{k}}/k_{\mathrm{B}}T}-1}=\sum_{n\mathbf{k}}\frac{A_{n\mathbf{k}}}{\mathrm{e}^{(\overline{\omega}+\delta_{n\mathbf{k}})/k_{\mathrm{B}}T}-1}\,.
\end{align}
The Einstein approximation assumes that it is possible to find a $\overline{\omega}$ such that the variations in the dispersion $\delta_{n\mathbf{k}}$ can be ignored, leading to $A=\sum_{n\mathbf{k}}A_{n\mathbf{k}}$. To go beyond the Einstein approximation, one can expand in small $\delta_{n\mathbf{k}}/\overline{\omega}\ll1$, 
\begin{widetext}
\begin{align}
\sum_{n\mathbf{k}}\frac{A_{n\mathbf{k}}}{\mathrm{e}^{\omega_{n\mathbf{k}}/k_{\mathrm{B}}T}-1}=\sum_{n\mathbf{k}}\frac{A_{n\mathbf{k}}}{\mathrm{e}^{\overline{\omega}/k_{\mathrm{B}}T}-1}-\sum_{n\mathbf{k}}\frac{A_{n\mathbf{k}}\mathrm{e}^{\overline{\omega}/k_{\mathrm{B}}T}\delta_{n\mathbf{k}}}{k_{\mathrm{B}}T(\mathrm{e}^{\overline{\omega}/k_{\mathrm{B}}T}-1)^2} 
+\sum_{n\mathbf{k}}\frac{A_{n\mathbf{k}}\mathrm{e}^{\overline{\omega}/k_{\mathrm{B}}T}(1+\mathrm{e}^{\overline{\omega}/k_{\mathrm{B}}T})\delta^2_{n\mathbf{k}}}{2(k_{\mathrm{B}}T)^2(\mathrm{e}^{\overline{\omega}/k_{\mathrm{B}}T}-1)^3}+\mathcal{O}(\delta_{n\mathbf{k}}^3)\,. \label{eq:method1}
\end{align}
\end{widetext}
This form provides a systematic way of improving upon the BE oscillator model, at the expense of increasing the number of fitting parameters. The even $\delta$ terms in the expansion are the most important ones because the $\delta$-spread about $\overline{\omega}$ is approximately equal on both sides, leading to a high degree of cancellations in the odd terms. This means that it is usually convenient to restrict the expansion to even terms.

\subsubsection{Two step method}

The phonon dispersion expansion introduces four fitting parameters, making it difficult to perform an accurate extrapolation with low-quality or sparse experimental data. We therefore propose an alternative method, based on fitting only the BE oscillator form to the experimental data, but requiring a recursive fit.

In general, the BE fit parameters $\mathbf{A}$ will depend on the maximum temperature included in the fit $\mathbf{A}=\mathbf{A}(T_{\mathrm{max}})$. As shown in Appendix\ \ref{app:extrapolation}, the high-temperature asymptotes for $\omega(T_{\mathrm{max}})$ and $A(T_{\mathrm{max}})$ in the BE oscillator fit are
\begin{align}
\omega(T_{\mathrm{max}})=& \,p_0+\frac{p_1p_0^2}{k_{\mathrm{B}}T_{\mathrm{max}}\ln(p_0/k_{\mathrm{B}}T_{\mathrm{max}})} \nonumber \\
&+\frac{p_2p_0^3}{(k_{\mathrm{B}}T_{\mathrm{max}})^2\ln(p_0/k_{\mathrm{B}}T_{\mathrm{max}})^2}\,, \label{eq:omega_t} \\
A(T_{\mathrm{max}})=&\, p_3\left(p_0+\frac{p_1p_0^2}{k_{\mathrm{B}}T_{\mathrm{max}}}+\frac{p_2p_0^3}{(k_{\mathrm{B}}T_{\mathrm{max}})^2}\right)\,. \label{eq:a_t}
\end{align}
This motivates a new scheme that can be implemented in two stages:
\begin{enumerate}
\item Fit the single BE, Eq.\ (\ref{eq:model}), to the data for a range of maximum temperatures $T_{\mathrm{max}}$.
\item Fit Eqs.\ (\ref{eq:omega_t}) and (\ref{eq:a_t}) to the functions $\omega(T_{\mathrm{max}})$ and $A(T_{\mathrm{max}})$ obtained in stage $1$.
\end{enumerate}
The final ZP correction is then $p_3p_0$.
This scheme only requires fitting of the two-parameter BE oscillator model to the experimental data.

%
%
%

\subsection{Benchmarking the extrapolation schemes} 

\begin{figure}
\centering
\includegraphics[scale=0.4]{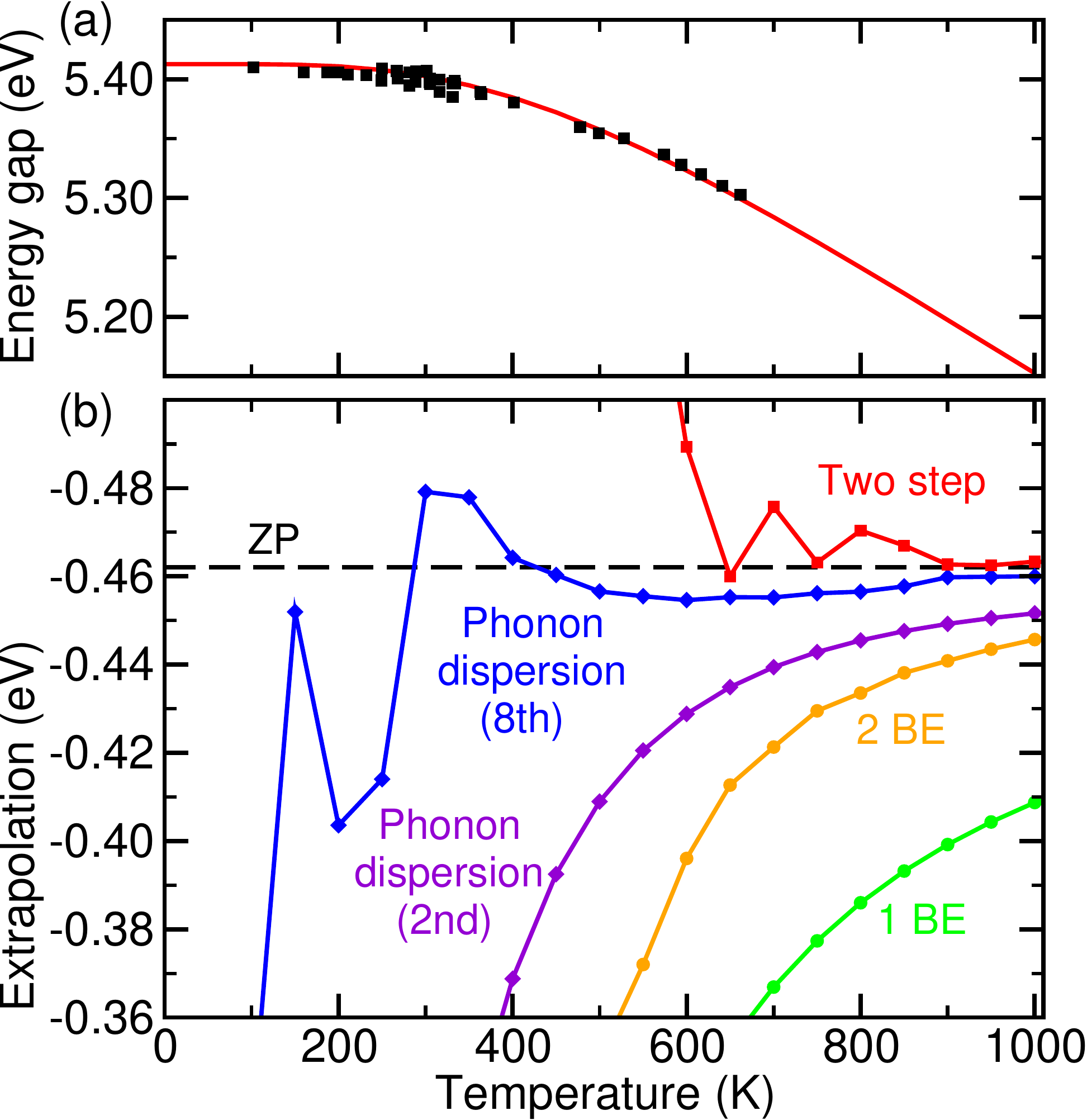}
\caption{(Color online) \textit{Upper}: Temperature dependence of the thermal band gap of diamond. The experimental data (black squares, from Ref.\ \onlinecite{Clark11021964}) are compared to the first-principles results (red line). \textit{Lower}: ZP correction to the thermal band gap of diamond obtained with the linear extrapolation scheme, using the most accurate models listed in Table\ \ref{tab:models}. } 
\label{fig:comparison_extrap}
\end{figure}

First-principles calculations of the temperature dependence of the thermal band gap of diamond\cite{PhysRevB.87.144302} provide a solid platform from which we can test the relative merits and accuracy of our two new models and the previous schemes. 
Diamond is a good case to study because both experimental data and first-principles results are available for the temperature dependence of the band gap. The upper part of Fig.\ \ref{fig:comparison_extrap} shows the temperature dependence of the band gap as given by Eq.\ (\ref{eq:bose}) including $162$ phonon modes (corresponding to a supercell with $54$ atoms) and with the couplings calculated from first principles using the method described in Ref.~\onlinecite{PhysRevB.87.144302}. The results of this calculation are in good agreement with experiment as shown in Fig.\ \ref{fig:comparison_extrap}, and the first-principles calculation gives a ZP correction to the gap of $-0.462$ eV. 

In Fig.\ \ref{fig:comparison_extrap} we show a comparison of the first-principles ZP correction and the extrapolated ZP correction using the fitting functions shown in Table\ \ref{tab:models}. The extrapolated ZP corrections from a fit to the first-principles data recover the first-principles ZP correction if data at sufficiently high temperatures is included. We have not shown data for the Varshni and P\"{a}ssler forms, which lead to poor results that only converge at higher temperatures above the range of the plot. 
Using one or two BE oscillators leads to reasonable fits allowing us to estimate the ZP correction. However, the convergence is slow, requiring data from temperatures of about $3,\!000$ K to estimate the ZP correction within $0.01$ eV. The P\"{a}ssler form has more degrees of freedom than a single BE oscillator and, even though (depending on the temperature range) it leads to a fit with a smaller mean square deviation, the extrapolation to zero temperature leads to worse results than fits based on the BE oscillator, and the extrapolated values are outside of the range of Fig.\ \ref{fig:comparison_extrap}. This can be explained by the emphasis of the P\"{a}ssler form on the low-temperature shape, which is not important for the asymptotic high-temperature limit or the ZP correction.

The new methods we have proposed lead to better estimates of the ZP correction. The phonon dispersion method with an expansion up to second order has the same number of fitting parameters as a double BE oscillator but consistently delivers better results. An expansion up to eighth order leads to results converged to better than $0.01$ eV above $350$ K. The two step method outperforms all but the phonon dispersion method with an eighth order expansion above $600$ K, and leads to results comparable to the latter above $900$ K.

\begin{table}[t]
\caption{ZP correction to the electronic band gap of diamond from the experimental data in Ref.\ \onlinecite{Clark11021964}.}
\label{tab:extrapolation_experiment}
\begin{tabular}{lc}
\hline
\hline
  & \hspace{0.1cm}  ZP correction (eV)   \\
\hline
BE oscillator & $-0.29$  \\
Phonon dispersion & $-0.41$ \\
Two step & $-0.51$  \\
Isotope (Ref.\ \onlinecite{Cardona20053}) & $-0.36$ \\
\hline
\hline
\end{tabular}
\end{table}

Now that we have established the limited applicability of the standard extrapolation methods and proven the accuracy of our two new methods, we are well-positioned to revisit the diamond experimental data discussed in Refs.\ \onlinecite{RevModPhys.77.1173,Cardona20053}. We use a variety of models to estimate the ZP correction, with the results summarized in Table\ \ref{tab:extrapolation_experiment}. The isotope method in Table\ \ref{tab:extrapolation_experiment} is an alternative approach for the determination of ZP band gap corrections, and it is described in Ref.\ \onlinecite{Cardona20053}. The single BE oscillator fit leads to poor results, in agreement with the theoretical assessment above.  We also note that the BE oscillator extrapolation value reported in Ref.\ \onlinecite{Cardona20053} disagrees with ours because we find different fit parameters than those reported there. The phonon dispersion result, using an expansion up to fourth order, leads to the better agreement with the estimate from the isotope effect, confirming it to be our recommended extrapolation tool. The two step technique does not perform as well as the phonon dispersion technique, as it is more sensitive to the absence of high temperature data, as seen in Fig.\ \ref{fig:comparison_extrap}.

Having completed the analysis of the experimental data, it is instructive to compare the estimate for the ZP correction to that from our first-principles calculations. The two new extrapolation schemes applied to the experimental data deliver $-0.41$ and $-0.51$~eV, lying in the same order as the theoretical assessment in Fig.\ \ref{fig:comparison_extrap}. This suggests an experimental ZP correction in the range $(-0.51,-0.41)$~eV, in good agreement with our first-principles result of $-0.46$~eV.

\section{Low temperature formalism} \label{sec:limits}

In recent years there has been a surge of interest in low-dimensional systems such as graphene and carbon nanotubes. When exploring the emergence of quantum critical physics at low temperatures it is important to understand the role played by phonons. 
In this section we extract the asymptotic behavior\cite{PhysRevLett.92.196403} at low temperatures from our framework, and extend it for the first time, as far as we are aware, to low dimensional systems. At low temperatures, only the lowest energy acoustic modes are excited, so these modes must be treated explicitly. Our derivation follows closely that in Ref.~\onlinecite{ashcroft}
for the specific heat. 

\subsection{Three-dimensional $T^4$ power law}

We first consider the three-dimensional system. 
In the limit of a large solid, the $\mathbf{k}$-points become dense on the length scale over which physical quantities vary appreciably. This allows us to replace summations over $\mathbf{k}$ by integrals over the first BZ of volume $V_{\mathrm{BZ}}$,
\begin{align}
\langle\hat{O}\rangle=\sum_n\int_{\mathrm{BZ}}\frac{\mathrm{d}^3\mathbf{k}}{(2\pi V_{\mathrm{BZ}})^3}\langle\hat{O}_n(\mathbf{k})\rangle\,.
\end{align}
The BE factor in the operator expectation value (see Eq.\ (\ref{eq:bose})) means that the occupancies of the modes with energies $\omega_n(\mathbf{k})\gg k_{\mathrm{B}}T$ vanish exponentially with decreasing temperature. This allows us to make four assumptions in evaluating the integral:
\begin{enumerate}
\item Only the three acoustic modes $\omega_n(\mathbf{k})=c_n(\hat{\mathbf{k}})|\mathbf{k}|$ 
contribute as $T\to0$.
\item The acoustic modes dominate within the BZ but vanish exponentially outside of it. We can therefore expand the range of the integral to the entire $\mathbf{k}$-space. 
\item As only the neighborhood of the $\Gamma$-point contributes to the integral, we can expand the couplings $a_{s;n}(\mathbf{k})$ in small $\omega_n(\mathbf{k})$, 
\begin{align}
a_{s;n}(\mathbf{k})=\sum_{p=2}^{\infty}a_{s;n}^{(p)}(c_n(\hat{\mathbf{k}})|\mathbf{k}|)^p\,. \label{eq:expansion1}
\end{align}
This expression uses the fact that $a_{s;n}(\mathbf{0})=0$ at $\mathbf{k}=\mathbf{0}$, corresponding to translational invariance. Also, the dominant term is quadratic (rather than linear) for a broad class of observables including the electronic band gap.\cite{PhysRevLett.92.196403}
\item The dominant term in Eq.\ (\ref{eq:full}) is the independent phonon term with $s=1$. 
\end{enumerate}
With these assumptions the expectation value reads
\begin{align}
\langle\hat{O}\rangle&=\sum_{p=2}^{\infty}\frac{3a_1^{(p)}}{2\pi^2V_{\mathrm{BZ}}^3c^3}(k_{\mathrm{B}}T)^{p+2}\int_0^{\infty}\mathrm{d}x\,x^{p+1}n_{\mathrm{B}}(x) \nonumber \\
&=\sum_{p=2}^{\infty}\frac{3\Gamma(p+2)\zeta(p+2)a_1^{(p)}}{2\pi^2V_{\mathrm{BZ}}^3c^3}(k_{\mathrm{B}}T)^{p+2}\,,
\end{align}
where $a_1^{(p)}c^{-3}=\frac{1}{3V_{\mathrm{BZ}}^3}\sum_na_{1;n}^{(p)}\int\frac{\mathrm{d}\Omega}{4\pi}c_n(\hat{\mathbf{k}})^{-3}$ and $\mathrm{d}\Omega$ is the infinitesimal angular element. $\Gamma(p+2)=(p+1)!$ is the gamma function for integer $p$, and $\zeta$ is the Riemann zeta function. In the low temperature limit the $p=2$ term dominates and we recover the power law of Ref.~\onlinecite{PhysRevLett.92.196403}, 
\begin{align}
\langle\hat{O}\rangle=\frac{\pi^2a_1^{(2)}}{10c^3}(k_{\mathrm{B}}T)^4\,,
\end{align}
with a prefactor that depends on the parameter $a_1^{(2)}/c^3$, which can be determined by fitting to experimental data.

We note that the third assumption above refers to the class of observables (including the band gap) for which $a_{s;n}(\mathbf{k})$ has a quadratic dependence on the energy around $\mathbf{k}=\mathbf{0}$.\cite{PhysRevLett.92.196403} However, the methodology presented here is more general, so it could be applied to other classes of observables with different asymptotic behaviour.


\subsection{Low dimensional systems}

In this section we extend the low temperature results to $2$- and $1$-dimensional systems. Low dimensional systems are important in understanding exotic physical properties and for technological applications, in particular graphene\cite{rise_of_graphene} and carbon nanotubes. Low dimensional systems are qualitatively different from $3$-dimensional systems because linear and quadratic acoustic branches coexist. This qualitatively different behaviour makes a detailed study of the low-dimensional systems essential.

In a $2$-dimensional system, there are two acoustic branches, one with linear dispersion, and the other with quadratic dispersion corresponding to out-of-plane atomic motion. 
 A $1$-dimensional system has a single acoustic linear branch and two quadratic acoustic branches. 

For $2$ and $1$-dimensional systems the linear branches lead to
\begin{align}
\langle\hat{O}\rangle_{\mathrm{2D}}^{\mathrm{linear}}&=\frac{2\zeta(3)a_1^{(2)}}{\pi c^2}(k_{\mathrm{B}}T)^3\,, \\
\langle\hat{O}\rangle_{\mathrm{1D}}^{\mathrm{linear}}&=\frac{\zeta(2)a_1^{(2)}}{2\pi c}(k_{\mathrm{B}}T)^2\,.
\end{align}
The quadratic branches with $\omega_n(\mathbf{k})=c_n(\hat{\mathbf{k}})k^2$ lead to
\begin{align}
\langle\hat{O}\rangle_{\mathrm{2D}}^{\mathrm{quadratic}}&=\frac{\pi a_1^{(2)}}{24 c^2}(k_{\mathrm{B}}T)^2\,, \\
\langle\hat{O}\rangle_{\mathrm{1D}}^{\mathrm{quadratic}}&=\frac{a_1^{(2)}}{8(\pi c)^{1/2}}\zeta\left(\frac{3}{2}\right)(k_{\mathrm{B}}T)^{3/2}\,.
\end{align}
The low-temperature asymptote of the lower-dimensional systems is dominated by the quadratic phonon branches. As far as we are aware, this is the first time these limits are reported and they are relevant for materials of reduced dimensionality such as graphene and carbon nanotubes.

A finite ($0$-dimensional) system has discrete phonon modes. Hence, the temperature dependence as $T\to0$ is a discrete sum of BE oscillators with an asymptotic exponential behaviour.


\section{Conclusions} \label{sec:conclusions}

We have presented an analytic phenomenology for describing the temperature dependence of phonon-renormalized properties. This formalism has allowed us to study important physical limits, and contextualize standard approximations and models used in the literature.

We first recovered from our formalism the usual BE expressions for the temperature dependence of electronic band gaps and lattice parameters. We have considered extensions to the high-temperature behavior beyond lowest order perturbation theory. We have studied standard extrapolation schemes for estimating the ZP correction of phonon-dependent properties from knowledge of their temperature dependence. The standard schemes fail to recover the correct asymptotic limit, and we have proposed and tested two new strategies that deliver results of an order of magnitude higher accuracy. Finally, we applied our new schemes to extract a more accurate value for the ZP correction for the band gap of diamond from experimental data.

We have also discussed the properties of the temperature dependence of band gaps in the limit $T\to0$. We have recovered the standard $T^4$ power law for three-dimensional solids, and obtained a $T^2$ power law in two dimensions and a $T^{3/2}$ power law in one dimension. These new results are important for materials of reduced dimensionality such as graphene and carbon nanotubes.

The theory we have presented makes no reference to a specific expectation value, microscopic theory, or material, and is therefore applicable to a wide range of phonon-related phenomena. The closed-form analytic results facilitate the calculation of further properties or limits beyond those explicitly considered in this paper.

We have treated the vibrational degrees of freedom within the harmonic approximation. To extend the formalism to include anharmonic properties, the anharmonic coupling between otherwise independent modes could be treated at mean-field level as in Ref.\ \onlinecite{PhysRevB.87.144302}. One could then expand the total wave function in terms of a simple harmonic oscillator basis, which would lead to matrix elements similar to those in Eq.\ (\ref{eq:matrix_element}). 

\begin{acknowledgments}
We thank Neil Drummond for useful discussions. 
B.M. and R.J.N. acknowledge the financial support of the Engineering and Physical Sciences Research Council (UK), and G.J.C. funding from Gonville and Caius College.
\end{acknowledgments}

\appendix

\section{BE oscillator derivation} \label{app:bose}

In this Appendix we present the derivation of Eq.\ (\ref{eq:bose}), the lowest order perturbation theory in terms of BE oscillators. We start 
by evaluating the sum over $p$ in the matrix element of Eq.\ (\ref{eq:matrix_element}) for $s=2$, 
\begin{align}
\sum_{p=\mathrm{max}(0,m-1)}^{m}\!\!\!\!\!\!\!\!\!({}^{m\!}C_{\!p})^{2}\frac{p!}{2^p(1\!-\!m\!+\!p)!}\! = \!\frac{m!}{2^m}\!+\!\frac{m^2(m\!-\!1)!}{2^{m-1}}\,.
\end{align}
We then obtain for the independent phonon term in Eq.\ (\ref{eq:full}), 
\begin{widetext}
\begin{align}
\sum_{m=0}^{\infty}\mathcal{M}_{2,m}\mathrm{e}^{-m\beta\omega_{n\mathbf{k}}}=\sum_{m=0}^{\infty}\frac{2!}{(4\omega_{n\mathbf{k}})^1}\left(1+\frac{m}{2}\right)(\mathrm{e}^{-\beta\omega_{n\mathbf{k}}})^m=\frac{2!}{(4\omega_{n\mathbf{k}})^1}\left(\frac{1}{2(1-\mathrm{e}^{-\beta\omega_{n\mathbf{k}}})}+\frac{\mathrm{e}^{-\beta\omega_{n\mathbf{k}}}}{(1-\mathrm{e}^{-\beta\omega_{n\mathbf{k}}})^2}\right)\,,
\end{align}
and finally
\begin{align}
(1-\mathrm{e}^{-\beta\omega_{n\mathbf{k}}})\frac{2!a_{2;n\mathbf{k}}}{(4\omega_{n\mathbf{k}})^1}\left(\frac{1}{2(1-\mathrm{e}^{-\beta\omega_{n\mathbf{k}}})}+\frac{\mathrm{e}^{-\beta\omega_{n\mathbf{k}}}}{(1-\mathrm{e}^{-\beta\omega_{n\mathbf{k}}})^2}\right)=\frac{a_{2;n\mathbf{k}}}{2\omega_{n\mathbf{k}}}[1+2n_{\mathrm{B}}(\omega_{n\mathbf{k}})]\,,
\end{align}
\end{widetext}
for a single phonon mode ($n,\mathbf{k}$). To obtain Eq.\ (\ref{eq:bose}) we then sum over ($n,\mathbf{k}$).

We can also evaluate terms with non-zero $a_s$ for $s\neq2$ in a similar fashion, for example, to obtain Eq.\ (\ref{eq:bose2}) with a non-zero $a_4$ term. The resulting sums over $m$ are then of the general form
\begin{align}
\sum_{m=0}^{\infty}m^rx^m=\mathrm{Li}_{-r}(x)\,,
\end{align}
where $x=\mathrm{e}^{-\beta\omega_{n\mathbf{k}}}$ and $\mathrm{Li}_{-r}$ is a polylogarithm of order $-r$. 

\section{Asymptotic temperature dependence} \label{app:extrapolation}

Here we present the derivation of Eqs.\ (\ref{eq:omega_t}) and (\ref{eq:a_t}). We perform a least-squares fit of the Bose-Einstein oscillator model to the data described by Eq.\ (\ref{eq:bose}). The square deviation of the model compared with the data is
\begin{align}
 \langle\Delta^2\rangle\!=\!\!\int_{\infty}^{\beta_{\mathrm{max}}}\!\!
 \Bigg(\underbrace{\!\frac{A}{\mathrm{e}^{\beta\omega}\!-\!1}\!-\!\sum_{n,\mathbf{k}}\frac{A_{n\mathbf{k}}}{\mathrm{e}^{\beta\omega_{n\mathbf{k}}}\!-\!1}\!}_{\text{$F(\beta,A,\omega)$}}
 \Bigg)^{2}
\!\!J(\beta)d\beta\,,
\end{align}
where we have retained the terms relevant for the temperature dependence, $J(\beta)=\beta^{-2}$ is the Jacobian, and $\beta_{\mathrm{max}}=1/k_{\mathrm{B}}T_{\mathrm{max}}$ is the maximum temperature of the data included in the fit. 

The fitting parameters ($A,\omega$) depend on $\beta_{\mathrm{max}}$. 
We take a small $\beta_{\mathrm{max}}$ (high temperature) expansion of Eqs.\ (\ref{eq:omega_t}) and (\ref{eq:a_t}), and the parameters ($A,\omega$) should obey
\begin{align}
\frac{\partial\langle\Delta^2\rangle}{\partial A}\! &=\!2 \!\!\int_{\infty}^{\beta_{\mathrm{max}}}\!\!\left(\frac{1}{\beta\omega}-\frac{1}{2}+\mathcal{O}(\beta)
\right)\!\!F(\beta,A,\omega)J(\beta)d\beta=0, \nonumber \\
\frac{\partial\langle\Delta^2\rangle}{\partial \omega}\! &=\!2A\!\! \int_{\infty}^{\beta_{\mathrm{max}}}\!\!\left(\frac{1}{\beta\omega^2}\!+\!\mathcal{O}(\beta)\!
\right)\!\!F(\beta,A,\omega)J(\beta)d\beta=0\,.
\end{align}
The $\beta$-dependence of the first term in the expansion is the same in both equations. For small $\beta$ we may neglect terms linear in $\beta$. 
After evaluating the integrals, we obtain two equations
\begin{widetext}
\begin{align}
&A\left[-\beta_{\mathrm{max}}+\frac{1}{\omega}\ln(\mathrm{e}^{\beta_{\mathrm{max}}\omega}-1)\right]-\left[-\beta_{\mathrm{max}}\sum_{n\mathbf{k}}A_{n\mathbf{k}}+\sum_{n\mathbf{k}}\frac{A_{n\mathbf{k}}}{\omega_{n\mathbf{k}}}\ln(\mathrm{e}^{\beta_{\mathrm{max}}\omega_{n\mathbf{k}}}-1)\right]=0 \\
&\frac{A}{\omega^2}\,\mathrm{Li}_2(\mathrm{e}^{-\beta_{\mathrm{max}}\omega})-\sum_{n\mathbf{k}}\frac{A_{n\mathbf{k}}}{\omega_{n\mathbf{k}}^2}\,\mathrm{Li}_2(\mathrm{e}^{-\beta_{\mathrm{max}}\omega_{n\mathbf{k}}})=0\,. 
\end{align}
\end{widetext}
This system of equations can be solved algebraically by means of the Newton-Raphson method in the high temperature limit to yield, to lowest order,
\begin{align}
\omega(\beta_{\mathrm{max}})&=\frac{g_{-1}}{g_{-2}} - \frac{6\beta_{\mathrm{max}}\ln{\frac{g_{-1}}{g_{-2}\Omega}}g_{-1}^2}{\pi^2\ln{\frac{\beta_{\mathrm{max}} g_{-1}}{g_{-2}}}g_{-2}^2}\,, \nonumber \\
A(\beta_{\mathrm{max}})&=\frac{g^2_{-1}}{g_{-2}}- \frac{6\beta_{\mathrm{max}}\ln{\frac{g_{-1}}{g_{-2}\Omega}}g_{-1}^3}{\pi^2g_{-2}^2},
\end{align}
where $g_{-1}=\sum_{n\mathbf{k}}A_{n\mathbf{k}}\omega_{n\mathbf{k}}^{-1}$, $g_{-2}=\sum_{n\mathbf{k}}A_{n\mathbf{k}}\omega_{n\mathbf{k}}^{-2}$ and $\ln\Omega=g_{-1}^{-1}\sum_{n\mathbf{k}}A_{n\mathbf{k}}\omega_{n\mathbf{k}}^{-1}\ln\omega_{n\mathbf{k}}$, and we have retained only the lowest order terms for each power. 
This recovers the lowest order terms for the $\omega$-dependence in Eq.\ (\ref{eq:omega_t}) and the $A$-dependence in Eq.\ (\ref{eq:a_t}). 

\bibliography{anharmonic}

\end{document}